\documentclass[review]{elsarticle}
\usepackage{lineno,hyperref,amsmath}
\usepackage{graphicx,subfig}
\usepackage{caption}
\usepackage{amssymb}
\usepackage{algorithm}
\usepackage{algorithmicx}
\usepackage{algpseudocode}
\usepackage[top=2cm, bottom=2cm, left=2cm, right=2cm]{geometry}
\usepackage{subeqnarray}
\usepackage{cases}
\usepackage{epstopdf}

\modulolinenumbers[5]

\journal{Journal of \LaTeX\ Templates}

\begin{document}
\nocite{*}

\begin{frontmatter}

\title{Differential cryptanalysis of image cipher using block-based scrambling and image filtering}


\author{Feng Yu, Xinhui Gong, Hanpeng Li, Xiaohong Zhao}

\author{Shihong Wang\corref{mycorrespondingauthor}}
\cortext[mycorrespondingauthor]{Corresponding author}
\ead{shwang@bupt.edu.cn}

\address
{School of Sciences, Beijing University of Posts and Telecommunications, Beijing 100876, China}

\begin{abstract}
Recently, an image encryption algorithm using block-based scrambling and image filtering has been proposed by Hua et al. In this paper, we analyze the security problems of the encryption algorithm in detail and break the encryption by a codebook attack. We construct an linear relation between plain-images and cipher-images by differential cryptanalysis. With this linear relation, we build a codebook containing $(M \times N + 1)$ pairs of plain-images and cipher-images, where $M\times N$ is the size of images. The proposed codebook attack indicates that the encryption scheme is insecure. To resist the codebook attack, an improved algorithm is proposed. Experimental results show that the improved algorithm not only inherits the merits of the original scheme, but also has stronger security against the differential cryptanalysis.
\end{abstract}

\begin{keyword}
\texttt{}Differential cryptanalysis\sep Codebook attack\sep Image filtering \sep Image encryption \sep Block-based scrambling

\MSC[2018] 00-01\sep  99-00
\end{keyword}

\end{frontmatter}

\linenumbers

\section{Introduction}

With the rapid development of computer networks and communication technology, protecting digital image transmission and storage in open network environment has become more and more important \cite{Katz12}. To cope with this problem, a number of image encryption algorithms have been proposed in recent years, especially chaos-based algorithms \cite{AdamsChaos1,EIChaos1,ChenChaos1,HuangChaos1,PareekChaos1,YoonChaos1}. Chaotic dynamics is suitable for cryptography due to its properties, such as pseudo randomness, ergodicity, sensitivity to initial values and control parameters. Since Matthews \cite{Matt16} proposed a chaos-based cipher, chaos-based cryptography has developed into a new branch of cryptography. Since Fridrich \cite{Fridrich7} first applied permutation-diffusion structure in design of image encryption, image encryption has been developing these years \cite{BelazPD2,AminaPD2,AhmadPD2,KhanPD2,PingPD2}. Chaos-based image encryption schemes utilizing various approaches have been proposed, such as improved diffusion \cite{ZhangD6}, variant keystream generation \cite{ChenKey7}, bit-level permutation \cite{CaoBL4,LiBL4} and plaintext-related permutation \cite{ZhangPR5,ChenPR5}.

However, some of the proposed schemes have been shown to be insecure. Li et al. point out that any permutation-only image encryption methods are unable to resist known/chosen plaintext attacks and only $O(log_L(MN))$ known/chosen plain-images can break the ciphers \cite{Li3}, where $M\times N$ is the size of images and $L$ the number of different pixel values. Li et al. break a diffusion-only image cipher with only one or two known plain-images \cite{Li2013}. Solak et al. analyze Fridrich's encryption algorithm by a chosen-ciphertext attack \cite{solak2010}, but the analysis method is useless while the algorithm has enough rounds of encryption.
In Ref. \cite{Chen24}, Chen et al. analyze a medical image encryption algorithm \cite{Fu22} by using differential cryptanalysis. $17$ chosen plain-images can reveal equivalent permutation key for 1-round and 2-round encryption. They propose a novel analysis method called double differential cryptanalysis comparison breaking multi-round encryption with $16N^{2}+1$ chosen plain-images, where $N^{2}$ is the size of the images. Basing on differential cryptanalysis, in Ref. \cite{Chen4} Chen et al. propose a codebook attack under chosen-ciphertext conditions and totally break multi-round cryptosystem \cite{Zhou23}.

Recently, Hua et al. propose an image cipher using block-based scrambling and image filtering (IC-BSIF) \cite{Hua1}, which is also permutation-diffusion type. It is well known that filtering is a common method in image processing and selecting appropriate filtering can deblur images. In Ref. \cite{Hua1}, IC-BSIF has been evaluated by all kinds of analysis methods. However, we find it insecure. We can construct a linear relation between plain-images and cipher-images by differential cryptanalysis and break the cryptosystem by a codebook attack.

The rest of the paper is organized as follows. Section 2 briefly describes the original image encryption algorithm. In Sect.3, we give some derivation of preparatory formulas. In Sect.4, we analyze IC-BSIF by using differential cryptanalysis and construct a linear relation between plain-images and cipher-images. In Sect.5, based on the linear relation, we propose a codebook attack and simulation results verify our theoretical analysis. In Sect.6, we give an improved approach to enhance the security and the last section summarizes the paper.

\section{Description of IC-BSIF}

The architecture of the original image encryption algorithm
IC-BSIF is shown in Fig.1. In Fig.1, $P$ and $C$ are
plain-image and its cipher-image, respectively. The encryption
process has four modules: block-based scrambling, image rotation,
image normalization and image filtering; $S^{(i)}$, $R^{(i)}$,
$N^{(i)}$ and $C^{(i)}$ are the output images of four
implementation modules, respectively, where $i$ is the $i$th
round, $i$=1,2,3,4. Sub-key $k^{(i)}$ generates a scrambling box
$O^{(i)}$ for block-based scrambling, a random matrix $Q^{(i)}$
for image normalization and a $3 \times 3$ matrix $M^{(i)}$ for
image filtering. In this paper, a uppercase letter stands for an
image or a matrix and a lowercase letter a pixel of the image or
an element of the matrix, for example, an plain-image $P$ and a
pixel value $p(x,y)$ at the position $(x,y)$.

To better understand cryptanalysis in Section 3, we will introduce
block-based scrambling, image rotation, image normalization and
image filtering in detail, and how to generate $O^{(i)}$,
$Q^{(i)}$ and $M^{(i)}$ refers to the original algorithm
\cite{Hua1}.

\begin{figure}[ht]

\centering
\includegraphics[width=10cm]{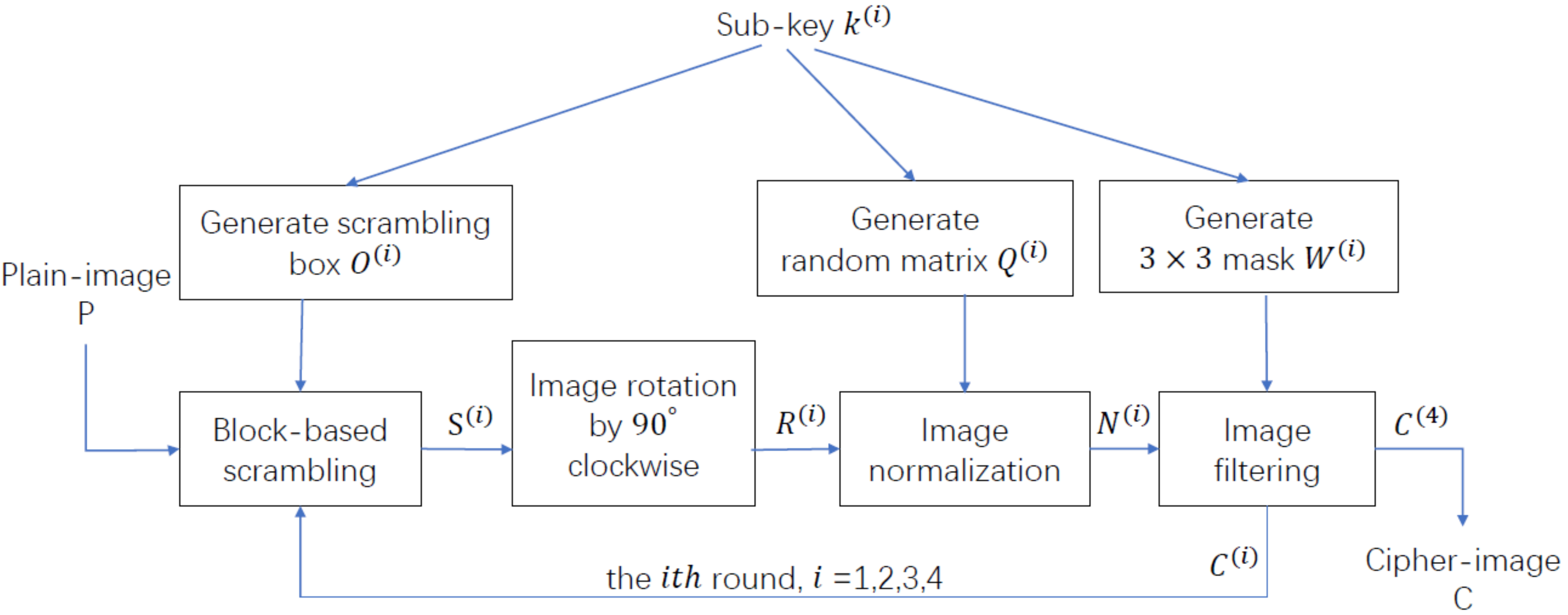}
\caption{The encryption process of IC-BSIF} \label {Fig.1}
\end{figure}

\textbf{Block-based scrambling}. This is a block-based permutation process and is
designed to weakness the strong correlation between the
neighboring pixels of plain-images. For an image of size $M \times
N$, the block size $L$ can be calculated by

\begin{equation}
L=min\{\lfloor\sqrt{M}\rfloor,\lfloor\sqrt{N}\rfloor\}
\end{equation}

The block-based scrambling is performed within range $L^2 \times
L^2$. The image of size $L^2\times L^2$ is divided into $L^2$
blocks and each block is of the size $L\times L$. All pixels in a
block can be permutated by using a scrambling Latin box $O^{(i)}$ of
size $L\times L$. In cryptanalysis, only permutation operation
cannot resist differential cryptanalysis.

\textbf{Image rotation}. For a plain-image of size $M\times N$,
because the block-based scrambling only shuffles its pixel
positions within range $L^2\times L^2$ randomly, the rest pixels
still locate at the unchanged positions. To shuffle all the pixel
position totally, the original algorithm takes image rotation by
$90^\circ$ clockwise after the block-based scrambling. Through four
rounds encryption, the image is rotated by $360^\circ$. In cryptanalysis, only rotation operation
cannot resist differential cryptanalysis.

\textbf{Image normalization}. The normalization operation to the rotated image $R^{(i)}$ using a random matrix $Q^{(i)}$ is defined as
\begin{equation}
n^{(i)}(x,y)=[r^{(i)}(x,y)+q^{(i)}(x,y)]\ mod\ F
\end{equation}
where $F$ denotes the grayscale level of the image. $F = 256$ if
the pixels of images are represented by 8 bits. In this paper, we
take $F = 256$.

\textbf{Image filtering}. The image filtering can change the pixel values randomly and
spread little change of the plain-image to the entire pixels of
the output image to achieve the diffusion effect. A mask matrix $W^{(i)}$ of size
$3\times 3$ is used to filter the normalized image. As shown in Fig.2, the 2-dimensional (2D) filtering operation of $n^{(i)}(x,y)$ is defined as
\begin{equation}
c^{(i)}(x,y)=[\sum_{j_{1},j_{2}\in \{1,2,3\}\cap(j_{1},j_{2})\neq(3,3)}w^{(i)}(j_{1},j_{2})c^{(i)}(x + j_{1}-3,y + j_{2}-3)+w^{(i)}(3,3)n^{(i)}(x,y)]\ mod\ F
\end{equation}
where $w^{(i)}(3,3)=1$ and other elements of $W^{(i)}$ are
produced by the sub-key $k^{(i)}$. The inverse operation of image
filtering is written as
\begin{equation}
n^{(i)}(x,y)=[c^{(i)}(x,y)-\sum_{j_{1},j_{2}\in \{1,2,3\}\cap(j_{1},j_{2})\neq(3,3)}w^{(i)}(j_{1},j_{2})c^{(i)}(x + j_{1}-3,y + j_{2}-3))]\ mod\ F.
\end{equation}

\begin{figure}[ht]
\centering
\includegraphics[width=8cm]{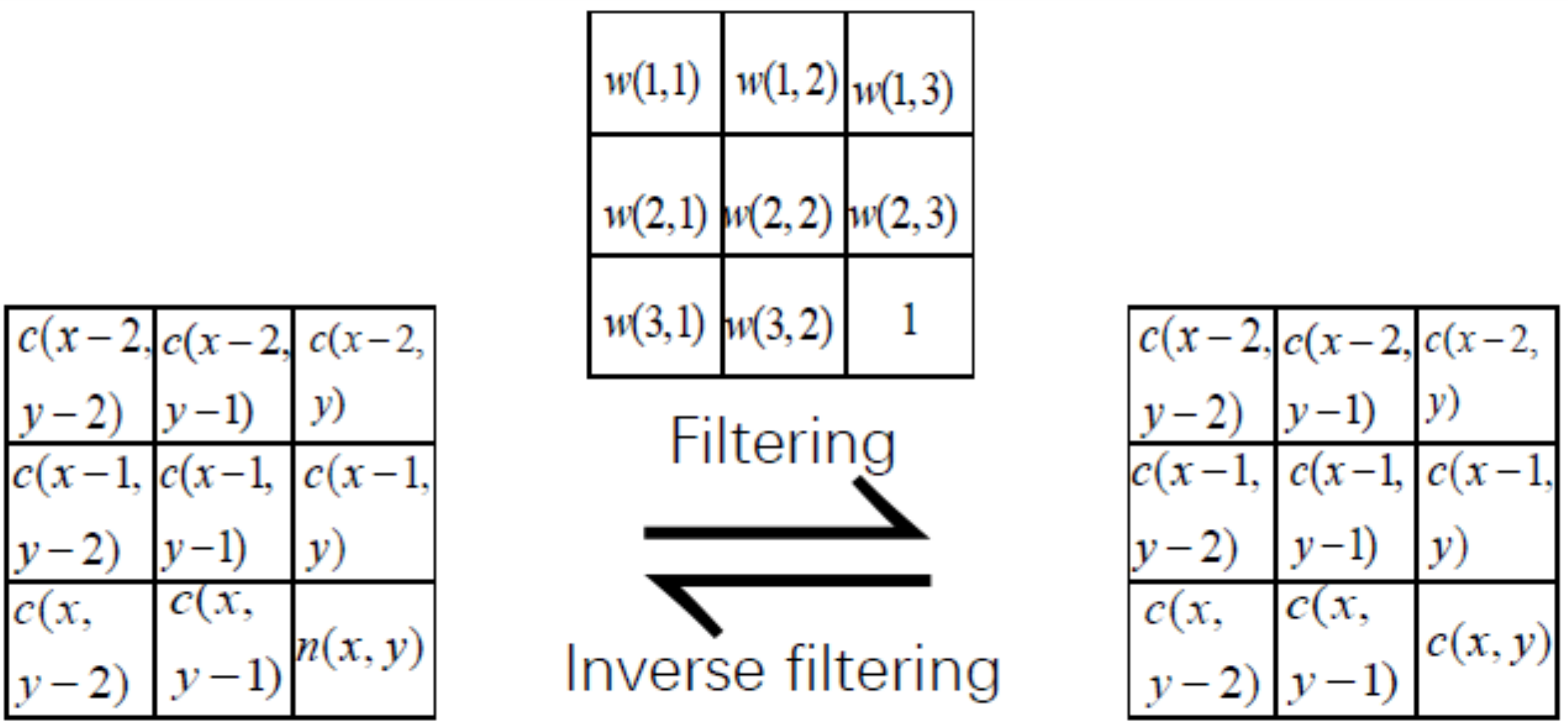}
\caption{An example of filtering operation and its inverse
operation. Omit all superscript symbols of Eqs. (3) and (4).}
\label {Fig.2}
\end{figure}

Here two points must be emphasized. (i) The upper and left adjacent
pixels are introduced to diffuse and confuse the current pixel
$n^{(i)}(x,y)$ in Fig.2. Therefore, the filtering operation begins
from the upper and left pixels of an image while its inverse
operation does from the lower and right pixels. (ii) For the
border pixels of an image in the leftmost column and the uppermost
row, filtering operations need two expanded columns and two
expanded rows, respectively. Taking the rightmost and lowermost
border pixels as the expanded border pixels in the leftmost column
and uppermost row, this strategy of expanding border pixels not
only masks all pixels, but also ensures the inverse filtering
operations.

\section{The preparatory work}


\textbf{Proposition 1}. Define $E(a_i)=(a_i + q)\ mod\ F$,
$i=0,1,2$. A differential equality is constructed by the following
expression $E((a_1 + a_2 - a_0) \ mod\ F )= (E(a_1)+E(a_2)-E(a_0))
\ mod\ F$.

\textbf{Proof}.
\begin{eqnarray}
E((a_1 + a_2 - a_0) \ mod\ F ) &=& (a_1 + a_2 - a_0 + q)\ mod\ F
\nonumber
\\
&=& ((a_1 + q)+ (a_2 + q)-(a_0 + q)) \ mod\ F \nonumber \\
&=& (E(a_1)+E(a_2)-E(a_0)) \ mod\ F \nonumber
\end{eqnarray}

\textbf{Proposition 2}. According to Propositon 1, a generally differential equality is constructed by the following expression $E((\sum^{n}_{i=1}a_i-(n-1)a_0) \ mod\ F )= (\sum^{n}_{i=1}E(a_i)-(n-1)E(a_0))
\ mod\ F$.

\textbf{Proof}.
\begin{eqnarray}
E((\sum^{n}_{i=1}a_i-(n-1)a_0) \ mod\ F) &=& (\sum^{n}_{i=1}a_i-(n-1)a_0 + q)\ mod\ F
\nonumber
\\
&=& (\sum^{n}_{i=1}(a_i + q)-(n-1)(a_0 + q)) \ mod\ F \nonumber \\
&=& (\sum^{n}_{i=1}E(a_i)-(n-1)E(a_0)) \ mod\ F \nonumber
\end{eqnarray}

\section{Differential cryptanalysis}

In this section, first we construct theoretically an linear
relation between plain-images and cipher-images for the original
algorithm by using differential cryptanalysis, then we give
simulation results of gray images.

\subsection{Differential cryptanalysis of one-round encryption algorithm}

First considering one-round encryption, we take three plain-images
$P_i$, $i=0,1,2$. The four specific operations, block-based
scrambling, image rotation, image normalization and image
filtering, have been defined as
 \begin{equation}
S^{(1)}_i=E_{s}(P^{(1)}_i),\ R^{(1)}_i=E_{r}(S^{(1)}_i), \
N^{(1)}_i=E_{n}(R^{(1)}_i), \ C^{(1)}_i=E_{f}(N^{(1)}_i), i=0,1,2.
\end{equation}

\textbf{Block-based scrambling}. Assume that the pixel at $(x,y)$
is mapped to $(x_1,y_1)$ through block-based scrambling, we write
this transformation by the following expression
\begin{equation}
E_s(p_i(x,y)) = s^{(1)}_i(x_1,y_1),i=0,1,2.
\end{equation}

We construct an input differential and calculate their output
differential expressed by the following forms
\begin{subequations}
\begin{numcases}{}
\Delta p(x,y) = [p_1(x,y) \pm p_2(x,y) ]\ mod\ F \\
\Delta s^{(1)}(x_1, y_1) = [s^{(1)}_1(x_1,y_1) \pm
s^{(1)}_2(x_1,y_1) ]\ mod\ F
\end{numcases}
\end{subequations}

Taking the differential $\Delta p(x,y)$ as the input, due to the
characteristic of permutation operation, we get the following
equality:
\begin{equation}
E_s(\Delta p(x,y)) =  \Delta s^{(1)}(x_1, y_1)
\end{equation}
Therefore, through block-based scrambling we construct the
differential equality of two images expressed by
\begin{equation}
E_s((P_1\pm P_2)\ mod\ F) = (S^{(1)}_1\pm S^{(1)}_2) \ mod\ F
\end{equation}

\textbf{Image rotation}. Same as block-based scrambling, given any
images, the rotation operation cannot change the relative position
of a fixed pixel of these images, thus we have the following
equality:
\begin{equation}
E_r((S^{(1)}_1\pm S^{(1)}_2) \ mod\ F) = (R^{(1)}_1\pm R^{(1)}_2)
\ mod\ F
\end{equation}
Combining Eqs.(9) and (10), we obtain the following differential
equality:
\begin{equation}
E_r(E_s((P_1\pm P_2)\ mod\ F)) = (R^{(1)}_1\pm R^{(1)}_2) \ mod\ F
\end{equation}

\textbf{Image normalization using the matrix $Q$}. Image
normalization of the original algorithm is the modular addition
operation of Eq.(2). Due to Proposition 1, we have the following
form
\begin{equation}
E_n((R^{(1)}_1+R^{(1)}_2-R^{(1)}_0) \ mod\ F) =
(N^{(1)}_1+N^{(1)}_2-N^{(1)}_0) \ mod\ F
\end{equation}
Especially, in Eq.(12) we choose three images as the input of
modular addition operation for eliminating the unknown matrix
$Q^{(1)}$ of Eq. (2).

\textbf{Image filtering}. By using differential cryptanalysis, the
filtering operation of Eq.(3) is transformed to
\begin{equation}
\Delta c^{(i)}(x,y)=[\sum_{j_{1},j_{2}\in
\{1,2,3\}\cap(j_{1},j_{2})\neq(3,3)}w^{(i)}(j_{1},j_{2})\Delta
c^{(i)}(x + j_{1}-3,y + j_{2}-3)+\Delta n^{(i)}(x,y)]\ mod\ F
\end{equation}
where $\Delta n^{(i)}(x,y)=n_1^{(i)}(x,y)-n_2^{(i)}(x,y)\ mod\ F$
and $\Delta c^{(i)}(x + j_{1}-3,y + j_{2}-3)=c_1^{(i)}(x +
j_{1}-3,y + j_{2}-3)-c_2^{(i)}(x + j_{1}-3,y + j_{2}-3)\ mod\ F $.
Based on the equation above, we can obtain $E_f(\Delta
n^{(i)}(1,1))=E_f(n_1^{(i)}(1,1))-E_f(n_2^{(i)}(1,1)) \ mod \ F$,
further acquire $E_f(\Delta
n^{(i)}(x,y))=E_f(n_1^{(i)}(x,y))-E_f(n_2^{(i)}(x,y)) \ mod \ F$,
$x=1,2,...,M$, $y=1,2,...,N$. Therefor for the filtering
operation, we have the following linear relationship:
\begin{equation}
E_f((N_1^{(1)}-N_2^{(1)})\ mod\ F) = (E_f(N_1^{(1)})-E_f(N_2^{(1)}))\
mod\ F
\end{equation}

Considering one-round encryption, through block-based scrambling,
image rotation, image normalization, and image filtering, we have
the following differential relationship for the three image
$P_0,P_1,P_2$
\begin{equation}
E_f(E_n(E_r(E_s((P_1+P_2-P_0) \ mod\ F)))) =
(C^{(1)}_1+C^{(1)}_2-C^{(1)}_0) \ mod\ F
\end{equation}

\subsection{Differential cryptanalysis of IC-BSIF and simulation experiments}
Although IC-BSIF undergoes four rounds of encryption, and the
sub-key of each round is different, our differential analysis of
Eq.(15) is still effective. Considering four-round encryption,
Eq.(15) is expanded to the following form
\begin{equation}
E((P_1+P_2-P_0) \ mod\ F) =(E(P_1)+E(P_2)-E(P_0))  \ mod\ F=
(C_1+C_2-C_0) \ mod\ F
\end{equation}
where $E$ denotes all the encryption operations of IC-BSIF.

Given any three plain-images, we can build Eq. (16) about the
plain-images and the corresponding cipher-images. Eq.(16) presents
a good linear relation. Next, we will verify Eq.(16) by using
simulation experiments.

Choose three plain-images of size $512\times512$, $Lena$ ,
$Baboon$ and a blank image. Encrypt $P_{lena}$, $P_{baboon}$ and
$P_{zero}$ by using the original algorithm, and obtain the
corresponding cipher-images $C_{lean}$, $C_{baboon}$ and
$C_{zero}$, shown in Fig. 3. To test Eq.(16), we calculate a
differential image $\Delta P=(P_{lena} + P_{baboon} - P_{zero})\
mod\ F$ and then encrypt it. The plain-image $\Delta P$ and its
cipher-image $\Delta C$ are shown in Fig. 3. We compute the
differential of the three cipher-images $\Delta C^{'}=(C_{lena} +
C_{baboon} - C_{zero})\ mod\ F$ and compare $\Delta C^{'}$ and
$\Delta C$. We find that $\Delta C^{'}=\Delta C$. The simulation
results confirm our theoretical analysis: Given any three
plain-images, we can build a differential relation between the
three plain-images and their cipher-images.

\bigskip
\begin{figure}[ht]

\centering
\includegraphics[width=10cm]{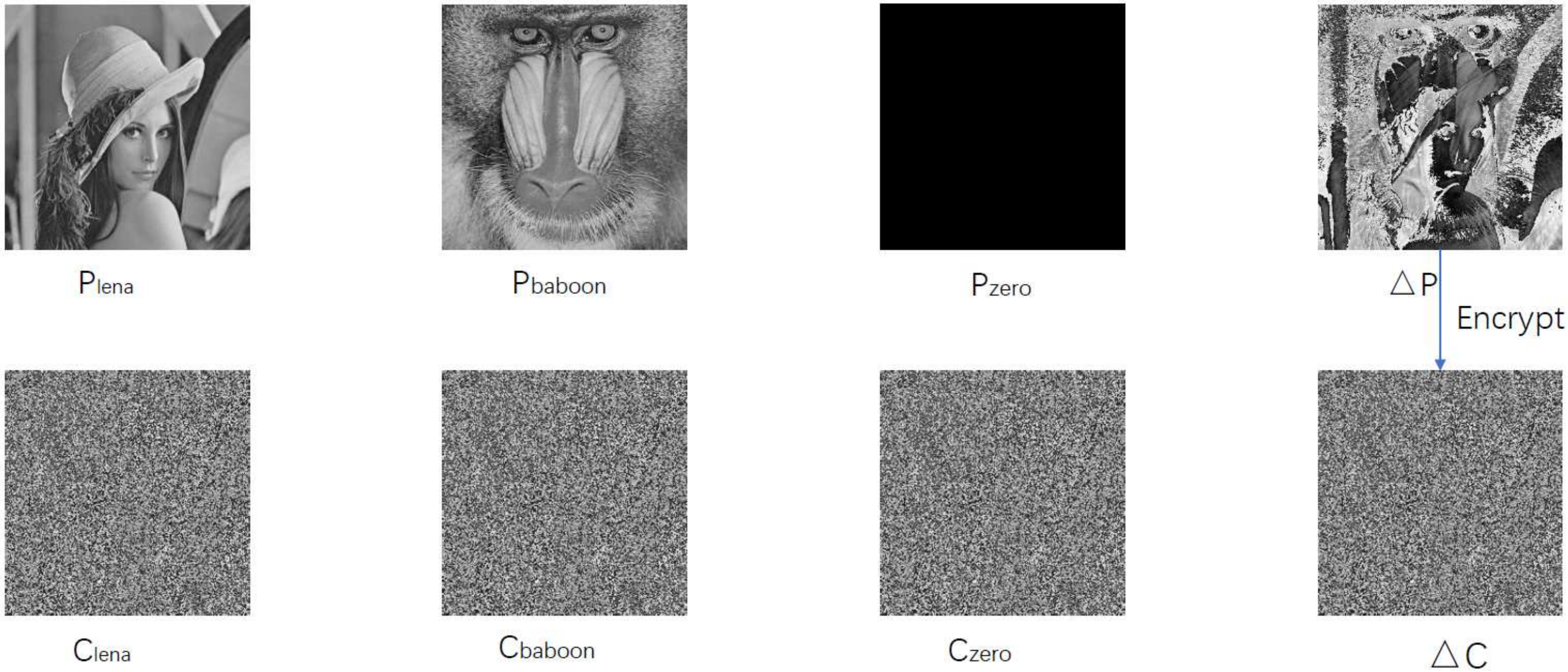}
\caption{Simulation results for testing Eq.(16)}
\label {Fig.3}
\end{figure}

 \section{Codebook attack}

 \subsection{Theoretical analysis}

 Select a blank cipher-image all pixel values of which are zero and $M\times N$ cipher-images with only a nonzero pixel $c_{(i-1)M+j}(i,j)=1$, $i=1,2,...,M,\ j=1,2,...,N$. Through decryption machine, we get the corresponding pairs of cipher-image/plain-image, i.e., $C_{0} \ and \ P_{0}$, $C_{n} \ and \ P_{n}$, $n=1,2,...,M \times N$. These pairs of cipher-image/plain-image are used for building a codebook and recovering any plain-images.

For given any cipher-image $C$, we first transform $C$ into the following form
\begin{equation}
 C=\sum^{M\times N}_{n=1}k_n\cdot C_n
\end{equation}
 where $k_n\in[0,255]$.

 We further transform the above expression into the following form
\begin{equation}
 C=\sum^{M\times N}_{n=1}k_n\cdot C_n - (\sum^{M\times N}_{n=1}k_n-1)\cdot C_{0}
\end{equation}
 and based on the differential cryptanalysis of Eq.(16) and Proposition 2, we naturally recover the plain-image expressed by the form
\begin{equation}
 P=\sum^{M\times N}_{n=1}k_n\cdot P_n - (\sum^{M\times N}_{n=1}k_n-1)\cdot P_{0} \ \  mod \ \ F
\end{equation}

 \subsection{Simulation results}
 Without loss of generality, we choose a image size of $64\times64$ as an example to represent the codebook attack. Algorithms 1 and 2 are two pseudocodes illustrating how to build the codebook and recover the plain-image. In Fig.4, we observe the plain-image $Lena$, the corresponding cipher-image and the recovered image by using the codebook attack.

  \begin{algorithm}
        \caption{Construct the codebook}
        \label{alg1}
        \begin{algorithmic}[1]
            \Require image size $M\times N$, a blank image with all-zero pixels and $M\times N$ images with only a nonzero pixel value ``1".
            \Ensure $M\times N + 1$ decrypted images. Construct codebook $P_{cb}$ ($M\times N + 1$ pairs of cipher-image/plain-image).

            \State $C = zeros(M*N+1,M,N)$
            \For{$i = 1 \to M$}
                \For{$j = 1 \to N$}
                    \State $C(j+((i-1)*M),i,j) = 1$
                \EndFor
            \EndFor
            \For{$i = 1 \to M\times N + 1$}
                \State $P(i, :, :) = decypt(C(i, :, :))$
            \EndFor
        \end{algorithmic}
    \end{algorithm}

    \begin{algorithm}
        \caption{Codebook attack}
        \label{alg2}
        \begin{algorithmic}[1]
            \Require cipher-image $C$, the grayscale $F$, and codebook $P_{cb}$
            \Ensure plain-image $P$
            \State $tmpImg = zeros(M,N)$
            \State $num = 0$
            \For{$i = 1 \to M$}
                \For{$j = 1 \to N$}
                    \If {$C(i,j)\!=0$}
                        \State $tmpImg = tmpImg + C(i,j)*(   P_{cb}(  (i-1)*M+j,:,:)   )$
                        \State $num = num + C(i,j)$;
                    \EndIf
                \EndFor
            \EndFor
            \State $P=\ mod\ (tmpImg - (num-1)*P_{cb}(M*N+1,:,:),\ F)$
        \end{algorithmic}
    \end{algorithm}

 \begin{figure}[ht]

\centering
\includegraphics[width=8cm]{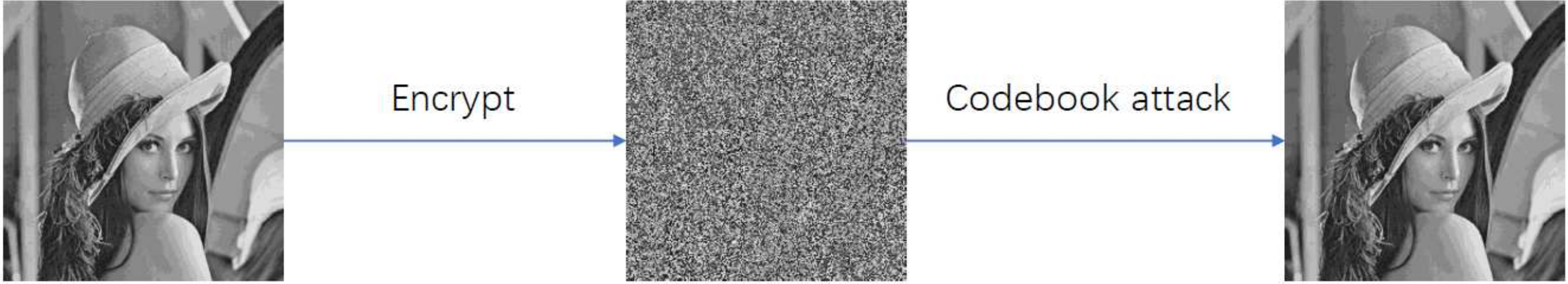}
\caption{Results of the codebook attack. $Lena$, its cipher-image and the recovered image by the codebook attack proposed.}
\label {Fig.4}
\end{figure}

\section{Improvements of IC-BSIF}
Using differential cryptanalysis, we construct a linear relation between plain-images and cipher-images for IC-BSIF. Based on Eq.(16), we can break IC-BSIF by the codebook attack. To resist differential cryptanalysis, we introduce image random rotations that are controlled by both plain-images and intermediate images in IC-BSIF. The improved approach is shown in Fig. 5. The block-based scrambling and image random rotation make up a group. In this group the scrambling and rotation operations have been alternately carried out four times. Then all modules including image normalization and image filtering are executed sequentially $m$ rounds, $m\geq 3$.

In the original algorithm, the image rotation is a regular rotation, i.e, 90 degrees clockwise. Through four times of rotation, the image is a return to original state. Here we propose image rotation controlled by a random index $(\alpha_1, \alpha_2, \alpha_3, \alpha_4)$. The angle of rotation of the $i$th time is equal to $(\alpha_i- \alpha_{i-1})\times 90$ degrees, where $\alpha_0=0$.  Rotate the image clockwise if the angle of rotation is greater than zero; otherwise, rotate counterclockwise. For example, $(\alpha_1, \alpha_2, \alpha_3, \alpha_4)=(2, 4, 3, 1)$. For the first rotation, the image is rotated by $180$ degrees clockwise; for the third time, rotate $90$ degrees counterclockwise. The random index is related to the plain-image $P$, intermediate images $C^{(i)}, i=1,2,...,m-1$ and the subkey $k^{(i)}$. We calculate the sum of pixel value of $C^{(i)}$, and take $(k^{(i)}\oplus 2^{\beta} \times sum(C^{(i-1)}))/2^{32}$ as the initial value of logistic map $x_0$, $i=1,2,...,m$, $C^{(0)}=P$, $\beta \in \mathbb{N}$ is set according to the size of images, here $\beta=0$. A chaotic sequence is produced by the form $x_{n+1}=4.0\times x_n\times (1-x_n)$, and by taking a segment with $n=101,102,103,104,$ and sorting the four variables a random index $(\alpha_1, \alpha_2, \alpha_3, \alpha_4)$ is generated.

In Figs. 6 and 7, we present the simulation results of the improved algorithm by using differential analysis. Same as in Fig.3, we encrypt four plain-images, $P_{lena}$, $P_{baboon}$,
$P_{zero}$ and $\Delta P = (P_{lena} + P_{baboon}-P_{zero}) \ mod \ F$ by using the improved algorithm, and obtain the
corresponding cipher-images $C_{lean}$, $C_{baboon}$,
$C_{zero}$ and $C_{\Delta P}$, respectively. We calculate the differential image $\Delta C=(C_{lena} + C_{baboon}-C_{zero}) \ mod \ F $, and $\Delta C^{'}=\Delta C-C_{\Delta P} \ mod \ F $, which are shown in Fig. 6. We observe that $\Delta C \neq C_{\Delta P}$ and $\Delta C^{'}$ illustrates random characteristic well. Compared with the original algorithm, the improved approach resist the differential cryptanalysis of Eq.(16). Because the image rotation controlled by a random index is introduced, the efficiency of the improved algorithm decreases about $20\%$ for four-round encryption ($m=4$) and increases about $20\%$ for three-round encryption ($m=3$).

Let's see the differential results for three-round encryption. Encrypt two images, ${lena}$ and a changed ${lena}$ with the last two pixels exchanged, and compute the differentials of cipher-images, $\Delta C^{(i)}=C_{Lena}^{(i)}-C_{changed}^{(i)}\ mod\ F$, $i=1,2,3$, shown in Fig. 7. The results demonstrate that the randomness of differential images increases while the number of encryption rounds increases and the improved system has a good randomness with $m\geq 3$.

We continue to do statistical test by National Institute of Standards an Technology (NIST) SP800-22 Statistical Test Suite \cite{Rukhin2015A,Pareschi2012On}. The significance level $\alpha$ is set as 0.01 and the number of binary sequences $s$ is set as 120. We choose 120 images from BOWS-2 image database and encrypt them by the improved algorithm. The cipher-images obtained are then decomposed into binary sequences. All the images are of size $512\times512$, thus the length of a binary sequence is $512\times512\times8=2097152$. The results show that 120 cipher-images encrypted by the improved algorithm can pass all the 15 sub-tests.

We also check the randomness of differential cipher-images. Same as in $Sect.4.2$, we encrypt the three plain-images using the improved algorithm, and obtain cipher-image $\Delta C$ and the differential image $\Delta C^{'}$. Then, we compute the differential image $\Delta C^{''}=\Delta C^{'}-\Delta C\ mod\ F$. The test results show that the differential cipher-image $\Delta C^{''}$ of the improved algorithm can pass all the 15 sub-tests, whereas $\Delta C^{''}$ of IC-BSIF does not due to $\Delta C^{'}=\Delta C$. This demonstrates that the improved approach can resist the differential cryptanalysis proposed by us.

\bigskip
\begin{figure}[ht]

\centering
\includegraphics[width=8cm]{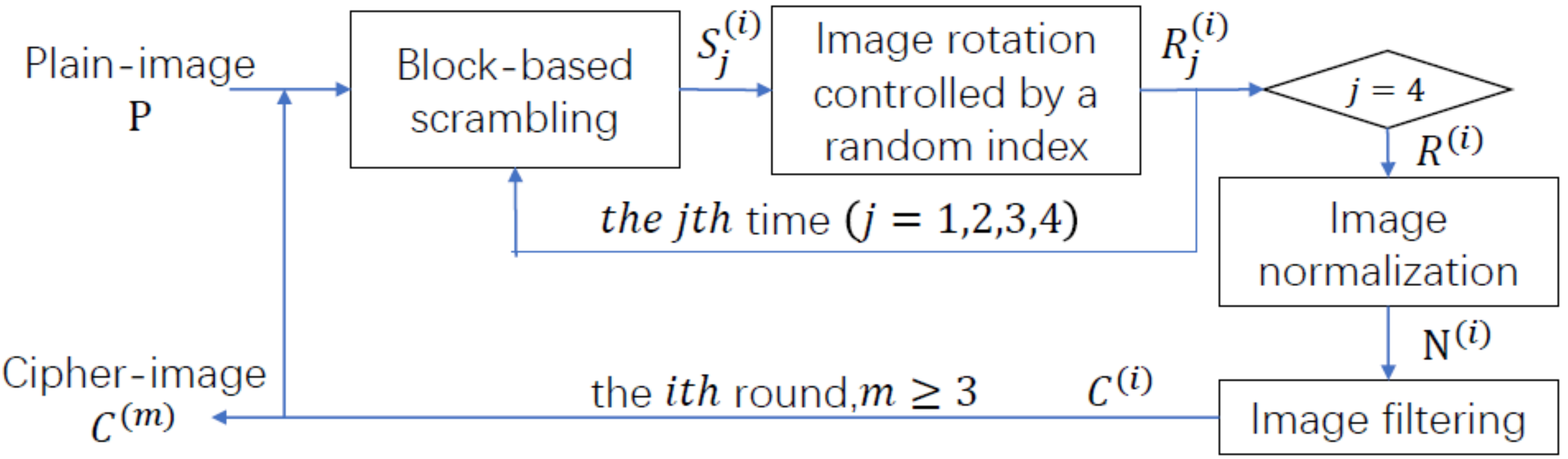}
\caption{The encryption process of the improved algorithm}
\label {Fig.5}
\end{figure}

\begin{figure}[ht]

\centering
\includegraphics[width=8cm]{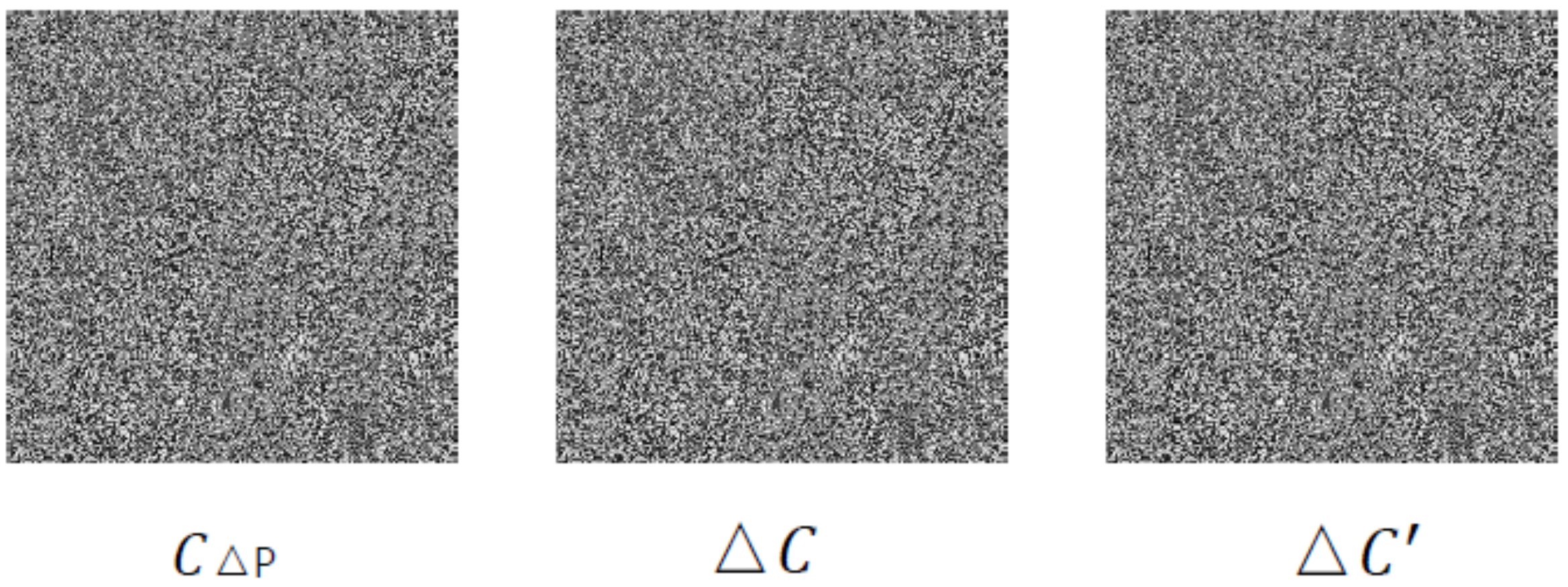}
\caption{Differential analysis of the improved algorithm for $P_{lena}$, $P_{baboon}$,
$P_{zero}$ and $\Delta P = (P_{lena} + P_{baboon}-P_{zero}) \ mod \ F$. (a) $C_{\Delta P}$. (b)$\Delta C$.  (c) $\Delta C^{'}=\Delta C-C_{\Delta P} \ mod \ F $.  } \label {Fig.6}
\end{figure}

\begin{figure}[ht]

\centering
\includegraphics[width=8cm]{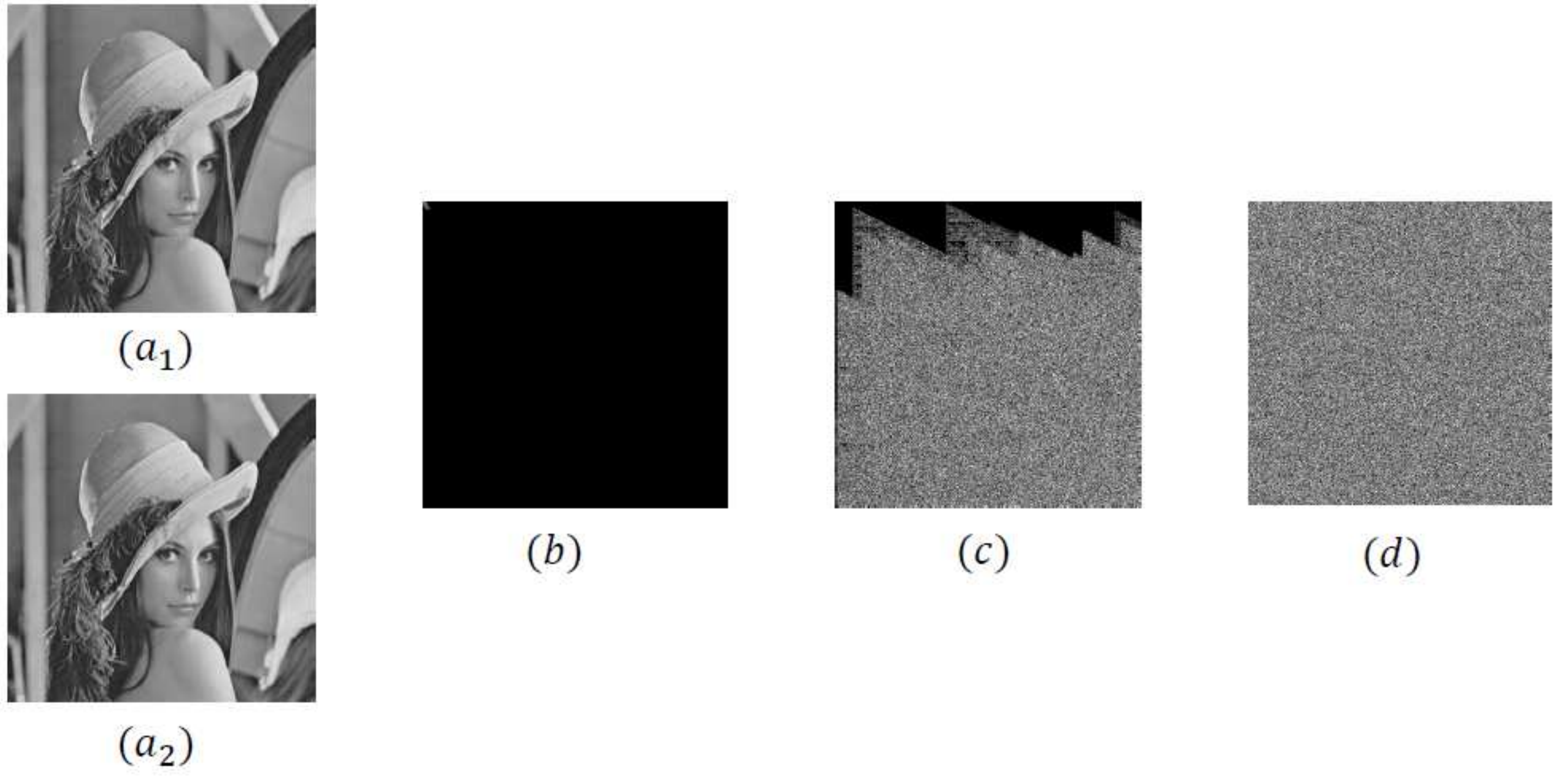}
\caption{Differential analysis of the improved algorithm for ${lena}$ (a) and the changed ${lena}$ (b) with the last two pixels exchanged. (b)-(d) The differential cipher-images of ${lena}$ and the changed ${lena}$ from 1 to 3 encryption rounds. } \label {Fig.7}
\end{figure}

 \section{Conclusion}

 This paper analyzes an image encryption algorithm using block-based scrambling and image filtering. We construct a linear relation between plain-images and cipher-images by differential cryptanalysis, although the encryption process is complex and nonlinear. Based on the linear relation, we build a codebook that contains $(M\times N + 1)$ pairs of plain-images and cipher-images, where $M\times N $ is the size of images. The proposed differential cryptanalysis and the codebook attack can be applied in analyzing a medical image encryption algorithm \cite{Hua2}. Enhancing the security of image encryption algorithms has been a challenge and we hope our analysis method will promote the research of image encryption to some extent.
\section*{References}

\bibliography{mybibfile}

\end{document}